\documentclass[manuscript, screen]{acmart} %review
\def\BibTeX{{\rm B\kern-.05em{\sc i\kern-.025em b}\kern-.08emT\kern-.1667em\lower.7ex\hbox{E}\kern-.125emX}}

\copyrightyear{2018}
\acmYear{2018}
\setcopyright{acmlicensed}
\acmConference[Woodstock '18]{Woodstock '18: ACM Symposium on Neural Gaze Detection}{June 03--05, 2018}{Woodstock, NY}
\acmBooktitle{Woodstock '18: ACM Symposium on Neural Gaze Detection, June 03--05, 2018, Woodstock, NY}
\acmPrice{15.00}
\acmDOI{10.1145/1122445.1122456}
\acmISBN{978-1-4503-9999-9/18/06}

\begin{document}

%
% The "title" command has an optional parameter, allowing the author to define a "short title" to be used in page headers.
\title[Caring for Alzheimer's Disease Caregivers]{Caring for Alzheimer's Disease Caregivers: A Qualitative Study Investigating Opportunities for Exergame Innovation}

%
% The "author" command and its associated commands are used to define the authors and their affiliations.
% Of note is the shared affiliation of the first two authors, and the "authornote" and "authornotemark" commands
% used to denote shared contribution to the research.
\author{Elizabeth Stowell}
\email{stowell.e@husky.neu.edu}
\affiliation{%
  \institution{Khoury College of Computer Sciences \& Bouve College of Health Sciences, Northeastern University}
  \streetaddress{360 Huntington Ave}
  \city{Boston}
  \country{USA}
}

\author{Yixuan Zhang}
\affiliation{%
  \institution{Khoury College of Computer Sciences \& Bouve College of Health Sciences, Northeastern University}
  \streetaddress{360 Huntington Ave}
  \city{Boston}
  \country{USA}}
\email{zhang.yixua@husky.neu.edu}

\author{Carmen Castaneda-Sceppa}
\affiliation{%
  \institution{Bouve College of Health Sciences, Northeastern University}
  \streetaddress{360 Huntington Ave}
  \city{Boston}
  \country{USA}
}

\author{Margie Lachman}
\affiliation{%
  \institution{Brandeis University}
  \streetaddress{415 South Street}
  \city{Waltham}
  \country{USA}}
  \email{lachman@brandeis.edu }

\author{Andrea G. Parker}
\affiliation{%
  \institution{Khoury College of Computer Sciences \& Bouve College of Health Sciences, Northeastern University}
  \streetaddress{360 Huntington Ave}
  \city{Boston}
  \country{USA}}
\email{a.parker@northeastern.edu}

%
% By default, the full list of authors will be used in the page headers. Often, this list is too long, and will overlap
% other information printed in the page headers. This command allows the author to define a more concise list
% of authors' names for this purpose.
\renewcommand{\shortauthors}{Stowell, et al.}

%
% The abstract is a short summary of the work to be presented in the article.
\begin{abstract}
The number of informal caregivers for family members with Alzheimer's Disease (AD) is rising dramatically in the United States. AD caregivers disproportionately experience numerous health problems and are often isolated with little support. An active lifestyle can help prevent and mitigate physical and psychological health concerns amongst AD caregivers. Research has demonstrated how pervasive exergames can encourage physical activity (PA) in the general population, yet little work has explored how these tools can address the significant PA barriers that AD caregivers face.  To identify opportunities for design, we conducted semi-structured interviews and participatory design sessions with 14 informal caregivers of family members with AD. Our findings characterize how becoming an AD caregiver profoundly impacts one's ability to be active, perspectives on being active, and the ways that exergames might best support this population. We discuss implications for design and how our findings challenge existing technological approaches to PA promotion.
\end{abstract}



%
% Keywords. The author(s) should pick words that accurately describe the work being
% presented. Separate the keywords with commas.

%
% A "teaser" image appears between the author and affiliation information and the body
% of the document, and typically spans the page.
%%  \includegraphics[width=\textwidth]{sampleteaser}
%%  \caption{Seattle Mariners at Spring Training, 2010.}
%%  \Description{Enjoying the baseball game from the third-base seats. Ichiro Suzuki preparing to bat.}

%
% This command processes the author and affiliation and title information and builds
% the first part of the formatted document.
\maketitle
\section{Introduction}
There are many forms of caregiving that are often undervalued and unseen within society. Caregivers of many kinds share similar burdens and experiences, such as prioritizing others' well being above their own, feelings of guilt for taking time for self-care, and care-related stress~\cite{gonyea2008adult, spillers2008family, ASPE2014c, chen2013caring}. However, caregivers are also a diverse population juggling both their own unique needs and the unique needs of those they care for. In particular, the number of informal caregivers who are caring for family members with Alzheimer's Disease (AD) is rising dramatically in the United States~\cite{ASPE2014c, ASPE2014a}, with the average length of care ranging from 8 to 10 years~\cite{farran2008lifestyle}. These informal caregivers are family members, partners, and friends of people with AD (i.e., the care recipients) who often live with the care recipient and provide much of the emotional and physical care required by the person with AD~\cite{Vitaliano2003, ASPE2014c}.  

AD caregivers disproportionately experience numerous physical and mental health problems, including insomnia, mood disorders, and stress associated with the burdens of care~\cite{Vitaliano2003, DeVugt2004}, to an even greater degree than non-dementia caregivers~\cite{suehs2013household, ASPE2014c}. Additionally, these caregivers are often isolated with chronic stress and little support~\cite{ASPE2014c, ASPE2014a, Boots2014, cheng2014benefit}. A key goal outlined in the recent US national plan to address AD is to enable informal caregivers to continue to provide quality care while maintaining their own health and well-being~\cite{ASPE2014b}. If a caregiver is able to successfully manage their own health and well-being needs, this can translate directly into more effective care for their AD care recipients~\cite{mittelman2006improving}.

There are a growing number of intervention programs designed to address AD caregivers' needs; however, these interventions are often focused directly on reducing stress and depressive symptoms~\cite{cheng2014benefit, dam2016systematic, Hirano2016,  Meichsner2016, Wennberg2015} and increasing self-efficacy~\cite{tang2016effects}. The Resources for Enhancing Alzheimer's Caregiver Health II (REACH II) program suggests that interventions not only focus on reducing negative problems associated with caregivers but ``also should focus on strengthening the positive experiences which may be equally important in enhancing health of family caregivers''~\cite{Basu2015}. One such positive health behavior is physical activity (PA), which has broad health benefits for cognitive, physical, and psychological health; however, caregivers face numerous barriers to PA, including the physical constraints of where caregiving can happen (e.g., the home, not at a gym), the unpredictability and frequency of care recipients needs, increase stress, isolation~\cite{farran2008lifestyle, rutman1996caregiving}. The barriers to PA that caregivers face are compounded by social isolation and limited social support for maintaining their own health and caring for the care recipient.

Games are defined as ``a type of play activity, conducted in the context of a pretended reality, in which the participants try to achieve at least one arbitrary, nontrivial goal by acting in accordance with rules''~\cite{adams2012game}. Within games, players expend effort in order to achieve a desired outcome~\cite{adams2012game}. Digital exergames, then, are software tools in which physical activity is required to drive game play and achieve the game objectives. A related concept is gamification, ``an informal umbrella term for the use of video game elements in non-gaming systems to improve user experience...and user engagement'' ~\cite{deterding2011gamification}. Much HCI and CSCW research has explored how the creation of exergames and the use of gamification can be meaningful ways of promoting physical activity~\cite{chao2015effects, larsen2013physical, saksono2015spaceship, Lin2006fishnsteps}. For simplicity, in this paper we will use the term ``exergames'' to refer to any kind of software tool that uses game mechanics (i.e., data, rules, and processes that determine how play progresses and how victory is achieved) or gamification to drive physical activity. 

Research has demonstrated the potential of exergame interventions to increase PA in a variety of populations~\cite{chao2015effects, larsen2013physical, saksono2015spaceship, Lin2006fishnsteps}; such platforms are promising means of encouraging exercise because they provide engaging media experiences through which people can set, pursue, and be rewarded for their PA goals.  However, less work has explored the potential of pervasive exergames for AD caregivers, that is, games in which play is not restricted to one setting (e.g., by a video game console) and in which a myriad of user activities may serve as input (e.g., step count while grocery shopping, vacuuming etc). Use of a mobile platform, such as a smartphone, enables the creation of more accessible interventions by removing locational and temporal constraints often present with console-based exergames~\cite{varshney2014mobile}. With pervasive exergames, the activities that people do throughout their day (and in varied settings) serve as input to the game experience~\cite{saksono2015spaceship, seifelnasr2015data}. These games have a great potential to more effectively address barriers to PA that caregivers face, such as the unpredictability and frequency of caregiving responsibilities, as they offer and encourage a wide range of PA experiences~\cite{wright2007persuasive, lieberman2009designing}. However, work is needed that characterizes how pervasive exergames can be made accessible, useful, and engaging for the AD caregiver population, given the significant challenges that they face.

Additionally, research has explored how technological innovations can support various caregiver populations with their caregiver responsibilities~\cite{hong2016care}, and the potential for technology-mediated social support amongst caregivers (e.g., through online support communities~\cite{tixier2010practices}). However, despite the fact that caregivers' needs are interdependent~\cite{chen2013caring}, prior work has typically explored the design of technologies that address discrete caregiver needs (e.g., self-care through PA or increased social connectedness). Therefore, research is needed to examine the specific barriers that AD caregivers face to wellness, and how these barriers can be addressed in concert. Our work seeks to address this research gap through a formative study exploring how pervasive games can jointly address barriers to PA and social connectedness in the AD caregiver population. Taking this approach allows us to explore the broader question of how technology can simultaneously address multiple caregiver needs; such an approach has the potential to be an effective and efficient means of enabling impactful engagements with health technologies in a population that is significantly overburdened. Our work is guided by the following research questions: 

\textbf{RQ1}: What are AD caregivers' attitudes towards PA (e.g., perceived barriers, benefits, and control) and toward social connection with other AD Caregivers? 

\textbf{RQ2}: What are AD caregivers' preferences for and attitudes towards a social exergame that encourages PA, social connectedness, and social support? 

We conducted semi-structured interviews and participatory design sessions with informal caregivers of family members with AD to understand the unique challenges around PA and wellness that AD caregivers face. Our findings characterize how becoming an AD caregiver profoundly impacts one's ability to be active, caregivers' perspectives on being active, and the ways that exergames might best support the rapidly growing AD caregiver population. We discuss implications for design, including the creation of social exergames that support caregiver's many roles and identities by celebrating time spent with the care recipient and supporting storytelling. We further discuss designing for strategic, socially-translucent connections, and leveraging caregivers' locational and temporal constraints.
\section{Background}
AD caregivers spend, on average, 8-10 years caring for their family member with AD~\cite{farran2008lifestyle}. Depending on the stage and symptoms of the person with AD, these caregivers can devote hours each day fulfilling tasks related to the care recipients self-care or mobility, household chores, transportation, and health or medical needs~\cite{ASPE2014c}. Because of the all-encompassing nature of their caring duties, AD caregivers often face higher rates of health issues such as stress, insomnia, and mood disorders than the general population or even than other caregivers~\cite{Vitaliano2003, DeVugt2004, suehs2013household}. These conditions not only impact the caregiver's quality of health, but they also impede a caregiver's ability to provide quality care to the family member.

Caregivers face a number of barriers to PA that complicate their access to existing health interventions.  These barriers include limited time for PA (and other wellness-promoting activities), limited social support from family and friends, and social isolation~\cite{farran2008lifestyle, ASPE2014c, ASPE2014a, Boots2014, cheng2014benefit}. These significant and compounding constraints necessitate novel technological applications that encourage PA by providing scaffolds that help caregivers surmount significant personal and interpersonal barriers to PA, social connections to motivate and model positive behaviors, and meaningful application interactions that caregivers feel confident incorporating into their daily routines. Our work is motivated by public health research demonstrating the importance of interventions promoting PA and social connectedness for promoting health and wellness among AD caregivers and those for whom they care.

\subsection{Physical Activity (PA) for AD Caregivers}
One important positive health behavior that has received little attention with regard to informal caregivers is physical exercise. There is compelling evidence that an active lifestyle has broad benefits for cognitive, physical, and psychological health~\cite{kohl2012pandemic, powell2011physical}, including many of the chronic conditions and the emotional symptoms found among caregivers, such as stress and depression ~\cite{Rodriguez-Sanchez2014}. In fact, ``the negative impact of caregiving on health is likely due, at least in part, to the reduced probability that caregivers engage in preventive health behaviors such as regular PA''~\cite{castro2002exercise}. 

Physical exercise should be considered as a useful treatment for caregivers, who could benefit from regular moderate-intensity exercise programs~\cite{orgeta2014does}. Caregivers are less likely to exercise than non-caregivers, especially given the time constraints associated with caregiving~\cite{castro2002exercise}. There have been a few interventions designed to increase PA~(e.g., walking) among AD caregivers~\cite{farran2008lifestyle, Rodriguez-Sanchez2014}. These have met with some success in terms of increasing activity and reducing depression, subjective burden, and stress~\cite{castro2002exercise}.

\subsection{Social Connectedness}
Caregivers are often socially isolated because of the time commitment and burden associated with taking care of the care recipient. In fact, the emotional and physical burden of care giving makes it challenging for caregivers to access and develop their social connections~\cite{palamaro2012effects}. Previous research on dementia caregivers has shown that limited social life among caregivers is one of the factors that lead the burnout~\cite{almberg1997caring}.

On the other hand, increasing meaningful social support from friends and family has been shown to help caregivers better deal with the stress and the strain associated with caregivers' life~\cite{yatchmenoff1998enrichment, chen2013caring, tixier2010practices}. For instance Song and Singer have shown that social support acts as a buffering effect on stress levels of caregivers of families with psychiatric disorders~\cite{song2006life}. Therefore, an in-home technology-based intervention that connects caregivers to each other and to their social circles has great promise as an effective platform for promoting wellbeing and health among AD caregivers.

\subsection{Use of Exergame Technology for AD Caregivers}
Outside of the caregiving context, researchers have extensively demonstrated the acceptability and efficacy of exergames for PA promotion in various populations~\cite{chao2015effects, larsen2013physical, saksono2015spaceship, Lin2006fishnsteps}. Exergames are digital games in which play is driven by and combined with real-world PA.  Much prior work has developed and tested digital coaching systems, whereby exercises are demonstrated in a stationary video gaming context (e.g., using the Nintendo Wii~\cite{chao2015effects}), the user's actions are tracked using computer vision techniques, and real-time feedback is provided regarding posture, form, and other elements of game performance~\cite{kurillo2015multi, van2013exergaming}. Such exergame interventions deployed in various populations have demonstrated widespread benefits, including improved strength, balance, flexibility, socialization, and positive mood~\cite{chao2015effects, larsen2013physical, unbehaun2018exploring}. However, there are stringent system requirements and other constraints, such as limited definition of exercises, that limit the applicability for widespread use within the AD caregiver population.

Unlike many of the exergames developed in the stationary video gaming context, pervasive exergame technology combines input from a wider variety and contexts of PA (e.g., step count during errands or movements during chores). 
Additionally, pervasive exergames provide opportunities for social interaction, structured guidance for PA and exercise, and a source of motivation and enjoyment~\cite{wright2007persuasive,lieberman2009designing}.  Since intrinsic motivation, enjoyment, and social support are key factors for sustained healthy behavior~\cite{lachman2010promoting, trost2002correlates}, exergaming may have the potential to decrease attrition and increase adherence to PA programs~\cite{rosenberg2010exergames}. Yet, developing effective digital games requires an iterative, participatory design process whereby the tool is tailored to the particular needs of the target population~\cite{seifelnasr2015data}. Tailoring PA exergaming for AD caregivers may contribute to improving PA participation and overall health and wellbeing in this population.

While research has demonstrated the potential of exergame interventions in a variety of populations~\cite{chao2015effects, larsen2013physical, saksono2015spaceship, Lin2006fishnsteps}, the AD caregiver population faces specific limits to PA that present new challenges for technology design. Providing care is time-consuming, which is one reason why caregivers have relatively low levels of PA~\cite{loi2014physical}. This suggests that technology-based exercise interventions, which can be accessed at any time and in any setting, would be particularly valuable and appropriate for this population. Indeed, a recent systematic review suggested that technological~(e.g., web-based) applications offer ``convenient, low-cost alternatives for delivering interventions to caregivers of persons with dementia''~\cite{godwin2013technology}. Incorporating exercise into a fun and enjoyable exergame that can be done at home has the potential to make regular exercise more feasible. In particular, limited work has involved caregivers in the design of technology-based system that directly support and encourage PA as well as social connectedness in the AD caregiver population~\cite{godwin2013technology, topo2009technology}. However, positive behaviors such as PA serve to improve the physical and emotional well-being of caregivers, a crucial step towards ensuring the effective care of AD patients.
\section{Method}
The focus of this work is to explore how pervasive exergames can address the various barriers to PA and social connectedness that informal caregivers  of family members with AD face. We conducted semi-structured interviews and participatory design sessions with 14 informal caregivers to understand the unique challenges to PA and wellness that AD caregivers face. 
Additionally, we used a participatory design process, in which caregivers provided input on their preferences for and attitudes towards pervasive, social exergames. This process yielded crucial input from our target population, resulting in identification of key factors that impact caregiver interaction with an exergame. 

\subsection{Participant Recruitment and Data Collection}
Participants were recruited through two partner organizations that serve caregivers of people with Alzheimer's disease, clinicaltrials.gov, local caregiver support groups, and craigslist. Caregivers were eligible for the study if they were 18 years or older, and if they were a full-time, at-home caregiver to a family member, partner, or friend with Alzheimer's Disease. Exclusion criteria was more than two errors on the Short Portable Mental Status Questionnaire~(SPMSQ), a standard to ensure that participants are cognitively healthy  ~\cite{pfeiffer1975short}.

Upon consent, we asked participants to fill out validated surveys related to PA and social connectedness. We measured exercise self-efficacy and control using a measure based on Bandura's model ~\cite{lachman2011relevance, neupert2009exercise}. We also measured social connectedness and social assurance ~\cite{lee1995measuring} and collected demographic data. Caregivers then participated in a semi-structured interview lasting about 30 minutes. The interviews probed daily routines, barriers, and supports for PA, as well as the caregivers' level of connection to a caregiver network, strong-ties (e.g., family members) and weak ties (e.g., neighborhood acquaintances). Participants were compensated \$50  for their time (\$25 for the interview and \$25 for the design activities). The Institutional Review Board at our universities approved the study protocol.

The interviews probed current barriers to and supports for PA, daily routines and the extent to which participants are (or are not) able to incorporate PA into their daily lives. {\textcolor{blue}{}PA was defined for participants as any movement of the body that uses energy, including exercise, household chores, running errands, or tasks while at work.} Interviews also explored participants' level of social connectedness to a caregiver network and other strong-ties (e.g., family members) and weak-ties (e.g., neighborhood acquaintances) ~\cite{granovetter1982othe}. Caregivers then engaged in a participatory design session lasting about one hour, during which participants engaged in card sorting and storyboard activities. Interviews and participatory design sessions were held in locations most convenient for our participants, including our University, local libraries, and participant homes. For one participant living a different region of the United States, sessions were conducted on Google Hangouts.

\subsection{Analysis}
Descriptive statistics were generated to characterize survey trends. Interviews and design sessions were audio recorded and transcribed to facilitate a thematic analysis inspired by  grounded theory analysis ~\cite{strauss1990basics}. Using the Atlas.ti 7 software, two researchers inductively coded transcripts separately, labeling emergent phenomena in the data. To reach inter-rater agreement, the researchers met regularly to discuss codes, reconcile discrepancies, and iteratively cluster codes to create higher-level themes. 

\subsection{Participant Overview}
14 caregivers participated in the study. Demographics are presented in Table \ref{tab:demographic_survey}. Most of the caregivers were female ($n=10$) and the average age of the caregivers was 50 years old ($SD=13$, range 30-77). The average age of participants was 50 ($SD=22$, range 30-77); however participants felt 4 years younger ($SD=22$, range 24-90), a measure of subjective aging. The average of Body mass index~(BMI) was 30.6~($SD=7.6$). In addition to BMI, Participants' scores on exercise self-control ($Median=4.42$, IQR= .44) and family/friend affectual solidarity scale ($Median=1.4$, IQR= .53) suggest healthier and more socially connected caregivers than is reflected in the literature ~\cite{almberg1997caring}. This is perhaps a product of the recruitment strategies, as most participants were recruited through organizations that provide supports to AD caregivers, and the study protocol, which limited participation to caregivers who could step away from their caregiving responsibilities to meet for two hours.

Caregivers looked after their spouses or partners ($n=4$), parents ($n=7$) or other family members ($n=3$, great grandparent, grandparent, sister in-law). The average age of the care recipient was 78 years old ($SD=10$, range 56-91). One of the 14 care recipients had early onset AD. Caregivers described filling various caregiver duties, with these responsibilities increasing in frequency and demand with the stage of AD. Caregivers described providing range of care, specific to the symptoms and comorbidities of the care recipient. For example, some care recipients were bedbound requiring assistance with mobility; however, all caregivers provided 24 hour care. 

\begin{table}
    \caption{Demographic Information. This table overviews our participants, including their age (and the age they feel), their gender, their body mass index (BMI), and their relationship to the care recipient.}~\label{tab:demographic_survey}
  \centering
  \begin{tabular}{l r r r r }
    \toprule
    %& & \multicolumn{2}{c}{\small{\textbf{Test Conditions}}} \\
   % \cmidrule(r){3-4}
    {\small\textit{ID}}
    & {\small \textit{Age~(Feel-like age)}}
    & {\small \textit{Gender}}
    & {\small \textit{BMI}}
    & {\small \textit{Care Recipient}}\\
   % & {\small \textit{ExeSelfEff}} 
   % & {\small \textit{ExeControl}} \\
    \midrule
    P1 & 48 (29) & M & 36 & Parent  \\
    P2 & 32 (25) & M & 21 & Other \\
    P3 & 38 (25) & M & 41 & Parent\\
    P4 & 52 (40) & F & 35 & Parent\\
    P5 & 30 (24) & M & 21 & Other\\
    P6 & 54 (40)  & M & 28 & Parent\\
    P7 & 43 (26) & M & 36 & Other\\
    P8 & 53 (47) & F & 20 & Parent \\
    P9 & 48 (83)& M & 25 & Spouse \\
    P10 & 32 (50) & M & 31 & Spouse\\
    P11 & 38 (77) & M & 33 & Spouse\\
    P12 & 52 (50) & F & 27 & Spouse\\
    P13 & 38 (35) & M & 29 & Parent \\
    P14 & 52 (90) & F & 45 & Parent\\
    \bottomrule
  \end{tabular}
\end{table}
\section{Design Activities}
\subsection{Inspiration Book}
To catalyze participants' design thinking, we first grounded their understanding of what an exergame could entail as the ice-breaker activity. To this end, participants were given an ``Inspiration Book'', as shown in Figure~\ref{fig:insp_card} (A), prior to the participatory design session in which they were asked to describe any games that they enjoy playing. The Inspiration Book used the Mechanics, Dynamics, and Aesthetics (MDA) game design framework~\cite{Hunicke2004} to scaffold participants' reflections on game design elements that can facilitate enjoyment.

\subsection{Card Sorting: MDA Framework and PA Barriers}
\begin{figure*}
\centering
\includegraphics[width=1\columnwidth]{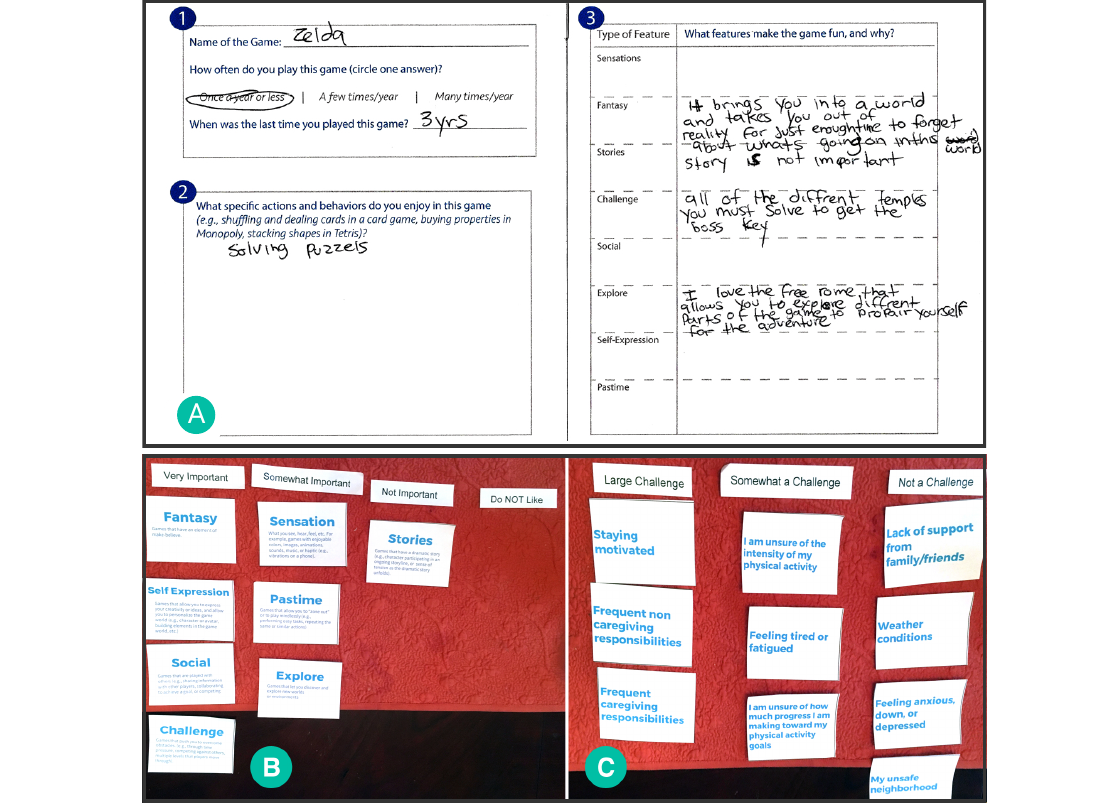} 
\caption{Design Activities. (A) Inspiration Book (B) Cardsorting: MDA Framework (C) Cardsorting: PA Barriers}~\label{fig:insp_card}
\end{figure*}

We used card sorting activities~\cite{hanington2012universal} to probe participant sentiment toward game design and toward PA barriers. In both card sorting activities, participants were asked to think aloud. Participants were asked to sort common barriers to PA that caregivers of family members with Alzheimer's often face~\cite{farran2008lifestyle}(Figure~\ref{fig:insp_card} (C)). Participants sorted the barriers into 3 rows, ranging from ``Large Challenge'' to ``Not a Challenge''. Participants were then asked to sort game aesthetics into 4 rows, ranging from ``Very important'' to ``Do NOT like''~\cite{Hunicke2004}~(Figure~\ref{fig:insp_card} (B)).

\subsection{Storyboards}
Participants then reviewed two storyboards that demonstrated various potential ways in which the exergame could be implemented for the AD caregiver population~\cite{hanington2012universal}. Given the time constraints of the caregiver, we opted to use pre-created storyboards as a springboard for their design thinking and to elicit feedback on contrasting concepts. Both storyboards presented exergames in which activity data was automatically collected by an activity tracking wristband, such as a FitBit~\cite{Fitbit} as well as user-provided input for activities that may not be captured by the FitBit (e.g., strength or flexibility exercises). The games depicted in the storyboards included four core design elements: 1) a self-paced instructional feature for learning PA strategies; 2) an experiential learning data dashboard that helps caregivers abstract insights from their data; 3) a digital rewards system, through which players earn virtual rewards individually and collectively with other caregivers; and 4) social engagement functions allowing caregivers to connect with and provide support to other users. The storyboards were designed to contrast types of PA motivations and social support, as well as to probe attitudes toward PA goal-setting, game themes, and barriers to PA and social connectedness.

\begin{figure*}
\centering
\includegraphics[width=1\columnwidth]{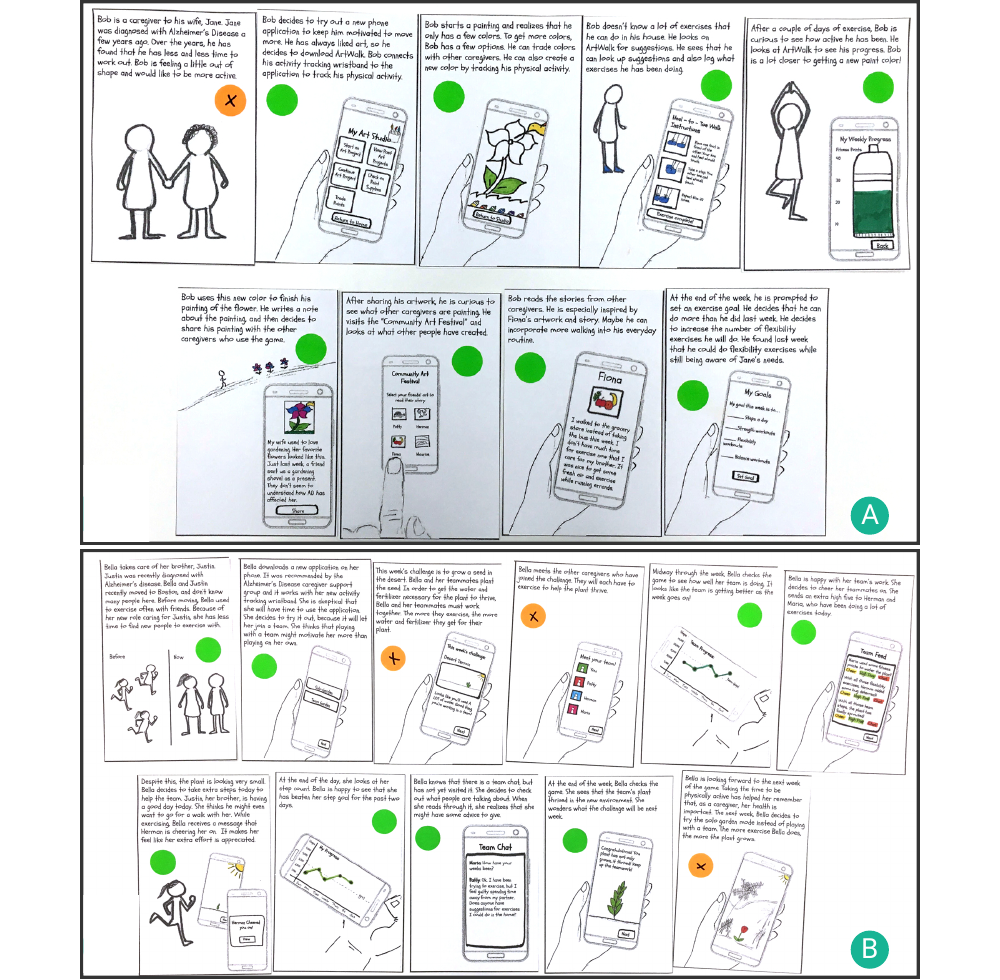} 
\caption{Storyboards. (A) ArtWalk (Top), (B) Garden-Themed Game.}~\label{fig:storyboards}
\end{figure*}

The first storyboard (ArtWalk), as shown in Figure~\ref{fig:storyboards}(A), described an art-festival themed game. This scenario presented a game played solo to examine the game mechanics of competition against the game and to isolate attitude toward teamwork in the context of PA and PA goals. The game contained elements of relaxation and stress management through the creation of artwork. Additionally, the storyboard explored the use of prompts and sharing stories as a means of promoting reflection and fostering a connection between caregivers. Caregivers were depicted responding to each others stories as a means of social support. The storyboard also demonstrated a caregiver referring to a database of exercises to choose from with instructions on how to perform the exercises. It also showed the caregiver setting  their own PA goals for the week and caregivers were prompted to think about how effective/comfortable they might be in setting their own PA goals. The storyboard featured a caregiver tracking their steps in a more abstract manner, with a tube of paint filling up as the caregiver took more steps.

The second storyboard (Garden), as shown in Figure~\ref{fig:storyboards}(B), described a gardening-themed game in which the caregiver protagonist played on a team with other caregivers towards a joint goal of growing a plant. This game contained elements of competitive play against the game and collaborative play with teams of caregivers combining their PA rewards to grow a flower. To elicit preferences toward team play, the final screen of the storyboard depicted a caregiver opting to forgo team play and to play alone. Additionally, this storyboard presented social interactions including a team chat and sending high fives or cheers to other caregivers. The caregivers in this storyboard were depicted engaging in involved forms of social interactions and social support, including teamwork, chatting, and checking in on PA progress. The storyboard featured a caregiver tracking their progress through a visualization that enabled them to see both their individual progress and the progress of their team.

A research assistant read the storyboards along with the participants. For each page of the storyboard, participants were encouraged to place an orange or green sticker on the page, as shown in Figure~\ref{fig:storyboards}, to indicate how much they did or did not relate to the experience of the caregiver in the storyboard. Participants were encouraged to think aloud as they moved from one page to another of the storyboard. We conducted semi-structured and unstructured interviews throughout the think aloud process. We conducted a semi-structured interview at the end of the each storyboard to collect initial impressions of the exergame presented. We also conducted a semi-structured interview at the end of the participatory design session to further explore participants' preferences and attitudes towards different exergame components.
\section{Findings}
Prior studies have shown that AD caregivers experience numerous barriers to PA and wellness, including caregiving and non-caregiving responsibilities, lack of support from family and friends, and poor mental and physical health, among others~\cite{farran2008lifestyle, Wennberg2015}. Each of these barriers is interconnected, and speak to the complexity of caregivers' lives. Our findings build upon this prior work by characterizing barriers experienced by our participants to PA and social connectedness, and how caregiver experiences shape PA and wellness motivations. We additionally detail caregiver perspectives on how technology might help them overcome these barriers. 

\subsection{Caregiver Constraints: Locational and Temporal}
Previous researchers have documented the demands of the caregiver role in various health contexts, including limited time to manage personal life, the all-consuming nature of the caregiving-role~\cite{kaziunas2015transition}, and unpredictable schedules~\cite{odonnell2000long,rutman1996caregiving, ASPE2014c, miller2016partners}; however, these experiences are compounded for AD caregivers, who spend on average 8-10 years as a caregiver, take on additional responsibilities as the the disease progresses, and must adapt to particularly unpredictable symptom progression~\cite{ASPE2014c}. In our study, we found that participants viewed their role as a caregiver as a barrier to PA because of a number of factors. These factors include the need to be physically present with the care recipient, the frequency of caregiver responsibilities, and the unpredictability of caregiving responsibilities. These factors impacted how AD caregivers felt they would interact with technological interventions. 

\subsubsection{Locational Constraints:} One of the major barriers that caregivers experienced to PA and wellness was the need to be physically present to fulfill caregiving responsibilities. 
Many caregivers described the extensive and constant care that care recipients require, and the need to be with the care recipient in the case that something go wrong. 
%The level of care required by the care recipients depends on the mobility and cognitive symptoms of the care recipient. 
%Depending on the level of care required by the care recipients, their mobility, and their cognitive symptoms,
Depending on the level of care required by the care recipient and the support the caregiver receives from family members and friends, this could result in the caregiver having very little privacy. 
%\textcolor{red}{The need to be physically present created two main barriers for caregivers 1) lack of access to PA facilities/spaces (e.g., gyms, going for runs/walks) / limitations for the physical spaces that the caregiver could visit and for how long and 2) lack of alone time to devote to wellness / a  constant ``audience'' when working out / lack of privacy.}
%\textit{``Because I don't want to leave her alone. She comes with me everywhere I go.''} (P4)
%\textit{``We're in the house a lot. We don't get out much. I've been told by the doctor not to leave him alone. So anyplace we go, we have to go together.} (P11)
Needing to be physically present to provide care resulted in caregivers being limited to locations or social contexts appropriate for providing care.
For many participants, this meant that much of their time was spent in the home. 
P11 described her need to be with her husband both to provide care and to be there in the case of an unexpected need:  
\begin{quote}
    I have to be there [in the house], and I make all of his meals... He can't make a phone call, which is one reason I really don't like to leave him... He couldn't call for help. He can't work the television. So as soon as I leave him, he'll press a button, and the television will be gone. And he loses things. Puts his clothes on backwards. He really can't do much on his own.
\end{quote}
P11 spends her day providing instrumental support to her husband for all of his self-care needs. 
P9 also discussed her inability to leave the house, saying: 
\begin{quote}
    ... if he's having a bad day, and he's tired, he'll literally sit in the same chair from after his milkshake, like 10:00 to 2:00, and I'm bored to death, but it's not like I can go anywhere or do anything.
\end{quote}
While P11 discusses how her husband's constant need for instrumental support limits her ability to leave the home, P9 focuses on the passive hours in which her main responsibility is monitoring her husband.
P9 does not have instrumental caregiving tasks to fill her hours in the home; however, she feels similarly anchored to her caregiving responsibilities. 
Both of the aforementioned examples illustrate situations in which the caregiver needs to spend time at home with the care recipient; however, our findings show that care recipients' ability to participate in social settings can also be another factor that limits the location in which caregivers' can be present.
P9 discusses a discouraging experience attempting to take her husband into a public setting.
\begin{quote}
    When [my husband] and I came out of the women's dressing room, and you're confronted with a female worker that says, ``Why are you in there?'' So, I try to be nice and like, ``Well, he has memory loss.'' And then she says, ``Well, you need to go in the mans side.'' And I say, ``Well, I'm the one trying on the clothes.''
\end{quote}
While P9's husband does not have issues with mobility, his need for constant monitoring creates uncomfortable encounters when out of the home. 
P2 similarly describes the challenges associated with bringing his grandmother in public resulting from others' misunderstandings of her condition:
\begin{quote}
    It is so difficult to bring her out, because she thinks she knows everyone. We were in the mall and we were taking the elevator and a family came in and they had a toddler and they had a stroller, a baby. And the next thing I know they're trying to get off. My grandmother has a death grip. Like her hands are just. She has this poor little kid. 
    He is screaming and crying, she won't let go. And I'm like, ``No, please, she's not trying to kidnap.'' I felt, and my mom was in the corner beet red. So we try to limit how much, how many people she sees in public. So we try to do things like where she... can't grab someone.% She means it in a nice way, but she just doesn't let go. Yeah, she just doesn't let go, so we have to be careful. (P2)
\end{quote}

These limitations also apply to social situations that are appropriate or comfortable for the caregiver and care recipient to engage in. 
P11 describes the rare social evening she is able to have with friends:
\begin{quote}
    He [husband, care recipient] can't seem to get out what he's thinking. And that's getting worse, noticeably, all the time. So when we're with other people, he doesn't say much at all because he can't express himself quickly enough to be part of the conversation. And sometimes he'll give this little talk about whatever, and nobody knows what he's talking about. I think he senses that. There's kind of this blank look. And it probably discourages him from speaking again.
\end{quote}
While this social event is not directly related to a PA, it reveals how care recipient ability to participate in social situations can limit caregivers' locality. For instance, a care recipients limitations for engaging in certain social situations could have implications for the contexts in which PA interventions for caregivers can be designed. 

%\textit{''...now if we're going and I'm bringing my mother and we're sitting down and we're doing something, she's only good for maybe an hour and a half, two hours. Next thing you know I'm trying to talk to you and she's trying to make her way for the door.''} P3

As care recipients develop more severe symptoms related to mobility or cognitive function, the area a caregiver is restricted to becomes smaller.
P7 describes this tightening locational constraint, explaining:
\begin{quote}
    Sometimes we'll go to the park. We haven't been doing that lately, because she [care recipient] hasn't been ambulating as well as before. As her disease is progressing she's just not moving as quickly or not moving at all.
\end{quote}
While walks in the park provided prior opportunity for PA for both the caregiver and the care recipient, this was no longer possible as the disease progressed. This progression requires new strategies for maintaining PA in the caregiver.
%P13, who lived with her mom after her mom had been diagnosed with AD, compared her experience of PA before becoming her mother's primary caregiver and after, saying:
%\begin{quote}
    %So right now my mom can't be left home alone at all and when she first moved in with me she was driving, she could be left home alone, she was cooking, she was doing all of her own bathroom stuff...
%\end{quote}

\subsubsection{Temporal Constraints:} 
We found that most participants viewed their responsibilities as a caregiver as a barrier to PA because of the lack of personal time resulting from frequent caregiving responsibilities. This theme emerged in the initial interview, and also through multiple design activities. In the card sorting activity, 12 out of 14 participants indicated that caregiving responsibilities were one of the most or somewhat challenging aspects about getting regular PA, as compared with other barriers~(e.g., neighborhood safety, caregiver health concerns, weather conditions). 

Additionally, we designed the first scene of the ArtWalk storyboard to examine how being a caregiver impacted participants' personal time and their PA motivation. 
The first scenario describes ``Bob is a caregiver to his wife, Jane. Jane was diagnosed with Alzheimer's Disease a few years ago. Over the years, he has found that he has less and less time to work out. Bob is feeling a little out of shape and would like to be more active.'' 
While a few participants felt they still have time to stay active (P2, P4, P8), most participants reported feeling the same as Bob in this scenario, indicating that they \textit{``have less time''} now that they are caregivers.
%Caregiving responsibilities introduced temporal challenges for our participants, as demonstrated by the results from our ArtWalk storyboard design activity. We designed the first scene of the ArtWalk storyboard to examine how being a caregiver impacted participants' personal time, and their PA motivation. The first scenario describes ``Bob is a caregiver to his wife, Jane. Jane was diagnosed with Alzheimer's Disease a few years ago. Over the years, he has found that he has less and less time to work out. Bob is feeling a little out of shape and would like to be more active.'' While a few participants felt they still have time to stay active (P2, P4, P8), most participants reported feeling the same as Bob in this scenario, indicating that they \textit{``have less time''} now that they are caregivers.

Our interview data further characterized the temporal constraints that our participants face. %For example, P3 expressed his frustrations to us: 
%\begin{quote}
%If I had somebody in here right now that could watch over her enough, three days a week, four days a week - and I could get out for an hour and a half just to go do my walks... I would go do that in an instant. (P3)   
%\end{quote}
While P3 currently has a few hours of support in the form of a personal care assistant who cares for his mother, he feels uncomfortable with the level of care that this assistant is able to provide. P3 explains, 
\begin{quote}
    [the personal care assistant] shouldn't be sitting on a chair looking at [her] cell phone while [my mother is] in the bathroom... you're supposed to make sure that [my mother]'s not doing what she's not supposed to do... [my mother] could get an infection.
\end{quote}
P3 expressed his deep concern over his mother's well-being and how these concerns put him ``on edge'' when he is away from her.
Even with help in the form of a personal care assistant, P3 indicates that the level of care provided in his absence is insufficient. This in turn means that it is difficult to make time for himself, because when he is away, he is in a near perpetual state of worry.
\begin{quote}
    She takes up the majority of my day. So when I have certain things to do it's very limited. Because I have to, I'm kind of like on a, like a stopwatch with her. You know, I hit the button, go. Come back, stop.
\end{quote}

%P13, who lived with her mom after her mom had been diagnosed with AD, compared her experience of being a caregiver before and after:

%\begin{quote}
   % So right now my mom can't be left home alone at all and when she first moved in with me she was driving, she could be left home alone, she was cooking, she was doing all of her own bathroom stuff ... (P13)
%\end{quote}
%She further explained how the role of being a caregiver impacted her PA. P13 was used to competing in triathlons and was on a swim team for much of her life. However, taking care of her mother consumed much of her free time, including her time for exercise and self-care. She recalled her experience with this barrier to PA: 
%\begin{quote}
    %I think before my mom lived with me I definitely went to the gym regularly... (P13)
%\end{quote}

%\begin{quote}
   % Something's gonna has to give if you have all this stuff to do and you know, you have to decide, ``I'm, I'm not going to get my workout. I'm going to have to sacrifice sleep or I'm going to have to not clean the house or I'm not.'' Something is going to have to slide in order for me to get the workout in. (P8)
%\end{quote}

\subsubsection{Unpredictable Schedule:} Another prominent theme was how the unpredictability of a caregiver's day impacts their attitudes toward technology that encourages PA. 
Our participants' challenges are grounded in a well-documented phenomena in the AD caregiving experience: uncertainty arises as caregivers do not know how the disease will progress~\cite{rutman1996caregiving, odonnell2000long}, and because they are unable to  control how the disease will impact the care recipient's behavior, emotions, and attention needs on any given day~\cite{rutman1996caregiving}. 
For example, P2 described his daily schedule usually depends on the care recipient: 
\begin{quote}
That can, that can vary depending on her state that day. One day she can be, you know, happy go lucky and just everything moves like a breeze... like yesterday for instance, it was one of the toughest days. Because she was just not having it. She was just very combative. She was just not in a good mood, you know.
\end{quote}

Health technologies that seek to encourage PA often attempt to help users build positive habits~\cite{Klasnja:2011:ETH:1978942.1979396}. 
Through goal-setting mechanisms and visualizations that emphasize and reward regular attainment of PA goals, routine and regularity are typically-embedded values within self-monitoring systems and pervasive exergames~\cite{Consolvo:2009:TDS:1518701.1518766, Klasnja:2011:ETH:1978942.1979396}. 
Yet, such embedded values are in tension with the reality of caregivers' lives. 
For example, P12 noted her inability to maintain a routine due to the unpredictability:
\begin{quote}
    I have tried changing my routine so that I get him through his morning rituals and then go meditate, and sometimes that works and often it doesn't, so I'm struggling with my own access to meditation time. That's one of my hardest things to manage.
\end{quote}
Despite the importance P12 places on meditation as a means of maintaining mental health during an emotionally trying time, she cannot rely on having this time to herself each day. 
Similarly, P6 expressed her difficulties with getting regular walk due to the caregiving responsibility:
\begin{quote}
    It depends on what you have for the day. Like today at this time I shall walk already. But I did not even start it today depending on what we have to do...
\end{quote}
P7 also described her caregiving routines, saying she has to \textit{``play it as it comes, because it's every day is different''}.
%``\textit{Then she’ll have those days. Like I said, I play it by ear as it, as it goes.}'' P7

%``\textit{It’s been kind of difficult lately, because as her disease is progressing, um, she doesn’t want to eat in the morning, so I have to wait a little bit later to feed her. So it [the day’s routine] all depends how that goes. If she eats in the morning then we’re good, we’re good to go, because you know we can do other stuff.}'' P7

\subsection{The Influence of the Caregiver Role on PA Motivation}
Becoming a caregiver for a family member impacts many aspects of an individual's life, including their motivations for PA.
For some of our participants, caring for a loved one with AD served as a warning, alerting them to the importance of taking preventative measures for healthy aging. Others expressed a need to take care of themselves because of their responsibility to the care recipient. Yet other caregivers did not associate change in PA motivation to their role as a caregiver.   

First, caring for a family member with AD served as a warning for some caregivers.
Assuming the role of a caregiver made some realize how important it is to develop healthy habits, because you see physical deterioration in front of you. 
P1 explicitly expressed how caring for someone with AD functions as a warning or motivation to establish healthy habits.
As caregivers become more familiar with people with AD, they understand \textit{``the physical agony''} that care recipients go through,  which motivates caregivers to be more active. 
P1 mentioned his top motivator for getting regular PA before he was a caregiver was \textit{``keeping in good health and weight''}. 
After caring for his mom and witnessing her physical agony, he became motivated to get regular exercise to avoid physical challenges as he, himself, ages.

Being a caregiver also made some caregivers realize the importance of changing their own habits because they have a responsibility to care for their loved one. 
For example, P7 and P12  noted that their responsibilities as a caregiver and the fact that their family member relies on them for care motivates them to exercise more. 
P12 described how the caregiving experience motivated her to do more exercises and stay healthy:
\begin{quote}
    One motivator right now is I can't take care of him if I don't take care of me. It makes it much, much clearer that I have to exercise, I have to sleep and eat and do all I can, because he depends completely on me.
\end{quote}

In some cases, PA was even tied directly to a caregivers ability to complete caregiver responsibilities. For example, P2 explained his experience of caring for his grandmother and how the caregiver responsibilities motivated him to do more Pilates to relieve back pain. 
\begin{quote}
    Giving her a shower is a challenge, because she can't stand and that's a lot of lifting. Pulling her up onto her wheelchair like if there's doctor's appointments of if she goes out or if I take her for a walk. That, one time my back did go out, because of that because I was pulling her up... That motivated me to do more Pilates, because I saw when I did the Pilates like my muscles. Everything was more aligned and I felt like my back felt a lot better. So the Pilates, it gives me more motivation to do Pilates when it comes to that.
\end{quote}

Similarly, P7 also believed not only the caregiving responsibility but also the emotional attachment motivated her to be more active. 
She explained: 
\begin{quote}
    I think it's just a sense of responsibility and that I have. I know that I have a really big responsibility that I, I have to do, because nobody else is gonna do it. That's what motivates me... She needs, she has to be cared for. And you know, even though she's my sister's mother-in-law. But I've known her for so long. I've known her for a really long time and we're attached, you know, by some way, shape, or form we are attached.
\end{quote}

However, we also observed that some participants~(P2, P5, P8, P9) believed that the caregiver role did not influence their motivation to exercise. 
One common characteristic shared by these participants is that they seem to have strong preexisting motivations to exercise. 
P2 discussed weight control as his motivation for getting exercise before being a caregiver.
\begin{quote}
    I want to feel healthy... I feel like I have a little gut. But no one says that I do, but I feel like I do... I don't want to gain weight. I don't want to gain weight.
\end{quote}
P2 said his motivations didn't change at all after being a caregiver. 
Similarly, P9 indicated that she usually cared about both being physically active and eating healthy even before becoming a caregiver. P9 compared the experiences of before and after being a caregiver: 
\begin{quote}
[\textit{Before}] Well, I always like being healthy. We ate healthy. No red meat. We always ate chicken, fish, fruit, vegetables. Although, we would have a cookie or cake, but we didn't eat greasy burgers or french fries with that. So, I have wanted to [be physically active]. I just can't unless he's distracted, or taken care of, or busy. However, you want to put it.

[\textit{After}] Feeling like garbage, but I'm so tired. I'm like, I have to do something with myself. I cannot sit here when you've sat so long that your arms lock up and there's only so many Andy Griffith reruns that a person can handle in a certain day. So, just like really not feeling well and knowing that I have to do something with myself. At the same time, feeling tired.
\end{quote}

%Interestingly, our survey results showed that these four participants were all within a healthy weight range, with BMI ranging from 20 to 25, and they reported a higher sense of exercise control and exercise self efficacy as compared to other participants. 
%Collectively, these findings show that the caregiving experience appeared to transform the PA motivations moreso for participants who seem to be already healthy. 

%P2 realized the physical benefits of being active, saying ``\textit{... because I saw when I did the Pilates like my muscles. Everything was more aligned and I felt like my back felt a lot better...}'' . 
%P8 brought up the mental benefits of doing PA as ``\textit{certainly mental health is huge. It just, because I am under a lot of stress and it really helps to center me and put things in perspective and calm me down and it’s just huge. It gives me more energy... It really delivers on all those fronts. Makes me a calmer, more balanced,}''

\subsubsection{Caregiver Guilt:} Some caregivers (P8, P9, P14) expressed their feelings of guilt about \textit{``taking care of myself''}. 
For example, P9 explained that 
    \begin{quote}
        We used to be able to play the Wii games that had a little bit more motion in them. Now, he just doesn't understand them anymore...Those were fun and he would be involved. Now, he's not so much. He's been limited. My big thing is, having something for him to do so I don't feel guilty taking care of myself.
    \end{quote}
The sense of guilt can become a barrier to doing exercises for caregivers because they \textit{``can't leave the house or they're really guilty or they're nervous leaving the house''}~(P8). 

At the same time, most caregivers~(P2, P6, P8, P9, P12) indicated that they would need time for self-care and also wanted to help other caregivers have more personal time. 
P9 mentioned she would love to write a blog on caregiving tips~(e.g., coping with new symptoms) to help other caregivers. 
However, she expressed her sense of guilt when doing something for herself or other caregivers, as P9 said that \textit{``then he'll just need something, or I'll feel guilty that I'm taking too much time away''}. 
Our findings regarding caregiver guilt are supported by prior studies~\cite{spillers2008family}, showing that caregivers reported a sense of guilt when they took time away from being a caregiver to fulfill their own needs. 
Caregivers realized they need to overcome feelings of guilt, and attempted various ways of creating time to address their own needs. 
The most common method of self-care was to recruit a family member or friend to take care of the care recipient for some hours. P9 described her experience entrusting her husband to a friend for a few hours so she can spend time on self-care:
\begin{quote}
    Okay, 30 minutes. Here's his water. Make sure he drinks. Don't leave him in the bathroom by himself... If they're willing to agree to those terms, then I'll let him go out. He did do that last week, twice, and he enjoyed it...
\end{quote}
However, this strategy only worked for short, sporadic periods of time, as the caregiver did not feel comfortable with her husband away from her.  
%Additionally, not all caregivers were able to count on this support from family and friends... this segues into the next section? 
%``\textit{it's my personality where I feel guilty that I don't like to do stuff without him [husband] or keep him busy because I don't like to see someone sort of, I call it melting. I don't want him to melt while I'm ignoring him.}''

\subsection{Technology-Supported PA and Social Connectedness}
Our participants reflected on ways in which technology could help them pursue PA while coping with the demands brought on by their role as a caregiver. 
Our findings show how PA promotion and nurturing social connectedness should go hand-in-hand for caregivers, but that this needs to be done carefully so as not to add additional burdens.
%[from the storyboard analysis: It is also important to note that prerequisite of goal-setting for people with a caregiver role is to incorporate caregivers' daily schedule into consideration, as P14 explained that the activities should not interrupt the care.]
Given the challenges inherent in caregiving, participants wanted to connect to other caregivers and saw the experiences represented in the storyboards as useful ways of supporting/promoting PA. 

However, they did not just want to connect to any caregivers, they indicated that there needs to be further commonality between them to develop trust. 

\subsubsection{Building Trust through Shared Traits \& Experiences:}
Another theme was the need to build trust in an exergame community through shared traits with the caregiver and the care recipient. 
While many caregivers felt that other caregivers would understand them, some felt that certain characteristics of the caregiver or care recipient would impact the extent of support they could receive. 
These characteristics included relationship to the person with AD, stage of AD, and symptoms of AD (given that there is a range of symptoms, each require different types and level of care). 
Some caregivers identified this themselves, noting that the types or extent of support they are able to get from other caregivers depended on certain shared characteristics. 
One participant described talking to an acquaintance who was her contemporary and who cared for her grandmother with AD:
\begin{quote}
    Talking to this woman [granddaughter caregiver, same age as P8] was really good and made me understand that it would be valuable to talk to somebody who's also a daughter. Like it's different. And it's different when you're 80 and when you're 53. It's a different dynamic. 
\end{quote}
P8 felt that her discussions with this caregiver acquaintance were validating in a way her local caregiver support group, made up mostly of spouse caregivers, was not. Her acquaintance had a shared experience of caring for a parent with little help from her siblings that the local support group could not directly relate to.

Another participant who cared for her husband with early-onset AD commented on how the stage and type of AD impacted the advice she felt she could receive from other caregivers, stating 
\begin{quote}
    Anything that I say, ``My husband does this and that.'' 
    With her, she's like, ``No. My husband doesn't do this, but he does this.'' I think it could be potentially dangerous or misleading to compare. Really with Alzheimer's, each person's situation is very unique.
\end{quote}
The fact that her husband has early-onset AD with less common symptoms meant that the informational support she received from others was often unhelpful. 
While the caregivers valued contact with other caregivers, they also felt that certain shared traits would provide them with more support.
%Some caregivers were wary of the support they could receive online.
%\begin{quote}
%... hopefully my team members or somebody within this app, you know, are positive enough. Because sometimes you get them people that are too cocky and they're like, ``Ah, well, yeah, you screwed up this time.'' 
%You know, and they say the mean words and the mean words aren't very good for someone that's trying very hard. (P3)  
%\end{quote}

\subsubsection{Support through Story-reading and Story-sharing:} 
Caregivers overwhelmingly were looking for informational support. 
When talking about experiences of connections with other caregivers, P2 thought ``\textit{that would benefit me if I talked to other people about how they handled certain situations like the sundowning.}''

The need for informational support was also identified from our storyboard activities. 
One scene from the ArtWalk storyboard was reading stories from other caregivers and sharing their own stories with them. 
\textit{``Bob reads the stories from other caregivers. He is especially inspired by Fiona's artwork and story. Maybe he can incorporate more walking into his everyday routine.''}
We found that participants showed their interests in story-sharing in different ways. 
Some caregivers (P4, P7, P11) seek caregiving advice and were more interested in borrowing lessons learned and coping strategies from other caregivers through story reading. 
Meanwhile, they would like to get inspiration for what other things they can do for the people they care for. 
There are also some participants (P8, P12, P14) who regard story-sharing as a bridge between other people who lack of understanding of the effects of AD on the both the care recipient and the caregiver. 
For example, P14 explained the situation in which family members have less understanding of her mother, the care recipient:
\begin{quote}
... because like with my mom, looking at her, if you talk to her, she would seem normal. Then when we're dealing with her, like to my son, oh it's just old age grandma, whatever. No, you don't understand. Take her for a day and you'll see what it's like.    
\end{quote}

P8 put green stickers to indicate she echoed with the scenario. She particularly advocated the story-sharing feature described in our storyboard since stories help people connect with each other. 
\begin{quote}
    I mean this is definitely a green. I guess I wish I could, I did more of this. This is like beautiful how he connects with the other caregivers and he shares something of himself... I think it's inspirational for other people using the app... And for them to understand that they, everybody, everyone is in the, everyone feels connected. They all have family members who don't understand what's happening with, with the, uh, person with Alzheimer's and yes, I definitely can put a green on that.
\end{quote}

\subsubsection{Building Relationships:}
Another recurring theme was the importance of building real relationships with people beyond virtual and online communication through games. While our participants collectively scored higher on the family/friend affectual solidarity scale, suggesting stronger social connectedness than has been described in previous caregiver literature, they still desired additional ways to access support and to better connect to the caregiver community.
A few participants (P5, P10, P12) expressed their desire to connect with people in physical realms. 
For example, P10 said that ``\textit{without being in actual contact with people, I don't really think that providing them much social support}''. 
Similarly, when talking about the motivation of playing exergames, P12 particularly highlighted the importance of being able to build real-world relationships. When asked how well she thought the games depicted in the storyboards could provide social support, she explained: 
\begin{quote}I think it depends on whether I end up with a relationship with at least one or two other people, as opposed to just anonymous names on a screen.
\end{quote}

Yet while caregivers wanted to build relationships, they also desired low-stress and low-demand interactions with and through the technology. For example, some caregivers expressed reluctance to playing team-based exergames. For example, when considering the game concepts in our storyboards, P4 told us:
\begin{quote}
    If I'm not in it [the team] it'd be fine. I'll cheer and I'll give high five. But if I have to be, nah, because they're [the team] gonna lose. 
\end{quote}
P4 felt that she would not be able to contribute effectively to a team-based exergame. While she saw the value in providing emotional support to others by cheering them on, actually playing together was not something she saw herself doing. P9 had similar concerns regarding team-based games, telling us:
\begin{quote}
    It could be encouraging. The person might want to do more that day to meet the goal, but at the same time, I wouldn't want to stress anybody out because I don't know what level their household situation is and I wouldn't ... Sometimes it would be a welcomed distraction and then sometimes it could be overwhelming. I could see the potential with team playing, to be overwhelming.
\end{quote}
P9's quote shows that while she sees the merit in team play, she felt that the dynamics of one's household demands could shift team play from being an encouraging experience to a daunting prospect. 

As we have discussed, caregivers face significant demands and high levels of responsibility as a result of their caring duties. Given these demands and the stress that they can introduce, our findings highlight that it is critical that technologies avoid introducing additional pressure into their lives. 

\section{Discussion}
Our findings characterize ways in which a caregiver's competing responsibilities and roles create a complex set of needs and barriers with respect to wellness and wellness promotion. Building upon prior CSCW and HCI research on caregiving~\cite{tixier2016counting, hong2016care, tixier2010practices, chen2013caring}, we use our findings to contribute new insight into the particular experiences of individuals caring for loved ones with AD, and caregivers' needs as they relate to computer-mediated social connectedness and physical activity support. An individual's role as an AD caregiver and the numerous responsibilities that come with it necessarily shape their priorities and ability to fulfill them. Being a caregiver complicates wellness promotion, but experiences as a caregiver can also provide motivation to be more active. On the one hand, caring for a loved one with AD meant that our caregivers were often locationally and temporally constrained as they needed to be physically present with their loved one much of, if not the whole day. The unpredictable progression of AD over time and even the unpredictable needs of a person with AD hour to hour or day to day, meant that developing PA routines is incredibly challenging. At the same time, for some, being a caregiver motivated them to be active, due to the visible demonstration of how one's body can decline or because they felt a sense of responsibility to be well enough to take care of their loved one. 

Our findings further characterize how participants felt that exergames might best support them in being physically active, including their perspectives on how such tools could most effectively connect them with other caregivers. We have focused our attention on social exergames in this work, given that prior work has demonstrated the promise of this technological medium for stimulating user engagement with physical activity behavior change and social interaction during this behavior change process~\cite{chao2015effects, larsen2013physical, Lin2006fishnsteps, saksono2015spaceship}. However, little is known about how exergames should be designed to provide benefits to the specific population of AD caregivers, who have unique challenges and who can benefit from increased supports for physical activity and social connectedness. Designing any type of health technology, including exergames, without investigating technological opportunities and limitations for AD caregivers risks the creation of health systems that are unusable for this population, which can further exacerbate the health disparities between caregiver and non-caregiver populations~\cite{veinot2018good}. Building upon the work of Veinot and colleagues~\cite{veinot2018good}, our research seeks to avoid the trap of intervention-generated inequalities, in which health systems designed without an eye towards the specific needs of vulnerable populations inadvertently create larger gaps in outcomes as they may be less accessible to and effective for these populations. Accordingly, our findings highlight several opportunities for how exergames can be specifically designed to encourage social connectedness and physical activity within the AD caregiver population. 

The complex set of barriers and needs experienced by our participants demonstrates the need for a more holistic approach to exergame design: one that addresses the multiple intersecting aspects of a caregiver's life. Building upon prior work that has called for caregiving tools that address caregivers' physical, emotional, social, and reflective needs~\cite{chen2013caring}, we provide recommendations for how such a multifaceted design approach can be explored within the domain of exergaming. Creating digital spaces where caregivers can simultaneously address many of their needs may promote wider technological adoption and use among an already overburdened population. Indeed, caregivers in our study expressed a preference for social components within a game, suggesting that the combination of multiple into an exergame needs (i.e., support for social connection and PA) would be welcome and even advantageous. 

In the following sections, we discuss considerations for future work focused on designing social exergames for the AD caregiver population. We show the unique considerations that must be made when designing for this demographic, including what existing approaches to exergame design show promise within an AD caregiver population, which need to be rethought to meet their needs, and open questions that require exploration through further empirical research. 

\subsection{Caregiver-Tailored Goal Setting}
An important aspect of behavioral change interventions is goal-setting (e.g., setting the goal of walking 7,000 steps each day this week)~\cite{shilts2004goal, locke1990theory, bandura1977role}. In particular, goal-setting is often a central component of exergame design, as successful progression in these games requires players to achieve behavioral change goals~\cite{saksono2015spaceship,Consolvo:2009:TDS:1518701.1518766, Lin2006fishnsteps, consolvo2008activity}. Prior work in health promotion more broadly has demonstrated the importance of grounding health intervention design in goal-setting theory, that is, frameworks that provide a way of conceptualizing how people can be supported in developing and assessing progress towards goals for change~\cite{Consolvo:2009:TDS:1518701.1518766, pearson2012goal}. In fact, prior work has shown that interventions grounded in goal-setting theories are more likely to yield positive health outcomes~\cite{shilts2004goal}. 

While researchers have examined goal setting in the context of health technology design broadly~\cite{consolvo2008activity,munson2012exploring,colineau2011motivating}, there is a paucity of work exploring specific goal-setting design considerations for technology that promotes wellness in caregiving populations. To address this gap in research, in this section, we use our findings to suggest important goal-setting considerations when designing exergames for AD caregivers. We specifically focus on the dimensions of goal difficulty and proximity, which are important for inducing self-efficacy for change, motivation, and ultimately, behavioral change~\cite{shilts2004goal, locke1990theory, bandura1977role, neubert1998value,pearson2012goal, strecher1995goal}. 

\textbf{\textit{Goal Difficulty:}} Prior health sciences research has demonstrated that behavioral change goals that are more difficult elicit more effort, as greater performance is needed to achieve them~\cite{shilts2004goal, locke1990theory}. Stated differently, when people pursue more challenging goals, they can be more likely to expend the effort needed to achieve those goals. Identifying the appropriate level of goal difficulty is an important part of any behavioral change intervention, given that goals that are too easy may not support individuals in reaching their full behavioral change potential, and goals that are too difficult may lead to reductions in self-efficacy, frustration, and disengagement from the behavior change process. 

Supporting goal-setting for AD caregivers requires even more nuance and care, because their perceptions of goal difficulty may vary day to day, and longitudinally as their loved one's disease progresses and their responsibilities evolve. Our caregivers discussed how their schedules were unpredictable due their uncertainty around how the disease will progress over time and how will impact their loved one day to day. Given this variability, their loved one might have a day in which they need less care thus opening up more opportunities for the caregiver to be active, whereas other days finding such time may be much more difficult. On easier days, a goal of 10,000 steps might seem achievable but on days when the person being cared for is experiencing significant challenges, this goal may be perceived of as unattainable.

The determination of what constitutes a ``difficult'' goal is likely to further depend on additional individual characteristics (e.g., personality, level of social support, additional life challenges and barriers to wellness, prior experience achieving behavior change goals, motivation, commitment to change, etc.). Research is needed that explores how exergames can support caregivers in creating personalized goals that are the appropriate difficulty level for them, given where they are in their caregiving journey. Digital tools that provide adaptive coaching support for goal determination---helping users navigate the complex and ever-evolving factors that may make change a challenge for caregivers---represent an important domain of exploration for this population. In addition, work is needed that determines best practices for goal variation, in terms of how frequently goal difficulty should be adapted to accommodate the constraints that caregivers face, versus holding goals constant and providing features that help caregivers to develop positive coping techniques that allow them to achieve their goals despite the challenges they face. 

\textbf{\textit{Goal Proximity:}} Another important aspect of goal-setting is how far the goal is projected into the future. Goals can be proximal (i.e., short-term) or they can be distal (i.e., long-term). For example, a proximal PA goal would be attempting to walk 10,000 steps tomorrow, whereas a distal goal would be to increase one's step count over the course of the month. Prior work has called for more research examining the optimal time windows for technology-assisted goal-setting, specifically research that investigates how time windows should vary ``based on the individual's activity level''~\cite{consolvo2009goal}. Our work extends this prior work by identifying the need for research that takes another approach---examining how the evolving experiences and needs of AD caregivers might impact the ideal proximity of their goals. 

For example, an important area of exploration is the extent to which the unpredictability and challenge of providing care necessitates tools that support more distal goal-setting or if more proximal goal-setting is ideal. Furthermore, while prior work in exergaming and health technology research more generally has enabled goal-setting as a core mechanism of the software experience, there is minimal evidence regarding how goal proximity should vary during different stages of one's life, and how to support such varied goal setting in a way that ultimately helps people move towards increased wellness. Determining how far into the future players will be allowed to set goals may significantly impact the design of the game experiences, given that game rewards are typically dependent upon goal attainment. As such, games based on user goals that are short-term (e.g., daily) may need to support different kinds of game mechanics (i.e., what the player is able to do within the game world) and reward structures (i.e., points, advantages, or other virtual benefits given to the player) than games based on longer-term goals (e.g., monthly). For example, an exergame with daily goals might provide smaller rewards than a game in which users pursue weekly or monthly goals. 

\subsection{Designing for Minimized Interactions}
Beyond goal-setting an important area of health technology research is nurturing app engagement. Prior work has demonstrated that health apps often have high drop-off rates, with users quickly abandoning them~\cite{druce2019maximizing, eysenbach2005law}. In response to this trend, one of the primary threads of research within health technology research has been exploring how to increase interaction with health systems. This goal has been sought in part because consistent engagement with the technology is seen as necessary to give users sufficient exposure to the technology to induce change. However, our work suggests that a different design orientation and research question is needed for the AD caregiver population, that is: what is the minimal level of engagement needed to support change? Framed differently, there is a need for work that examines the sweet spot of minimizing user interaction and maximizing benefits. 

Caregivers in our study expressed the all-encompassing nature of their role as a caregiver. Even when these caregivers were able to take time away from the care recipient, they described the guilt they felt in doing so and their fear that something would go wrong while they were away. Therefore, when creating exergames, designers should carefully consider how it may enhance and celebrate the individual's role as a caregiver without distracting from caregiving responsibilities. This requires consideration of the locational and temporal constraints that caregivers face. We suggest that such constrains should be directly leveraged as part of the game mechanics in future systems for this population. 

One path forward is exploring how exergame reward structures should be tailored to the caregiving population. Rewards are an intrinsic part of game design and app gamification, as they provide an incentive for progressing within the game and achieving the game objective~\cite{adams2014fundamentals}. In the context of exergames, goal achievement is typically framed in terms of the player accomplishing a PA goal~\cite{Lin2006fishnsteps, saksono2015spaceship, foster2010motivating, fujiki2008neat, macvean2012ifitquest, miller2014stepstream,xu2012not}. However, our work suggests that additional forms of achievement should be celebrated when designing exergames for caregivers. For example, to accommodate caregivers' locational constraints, an exergame could be designed to honor the value of spending time with the care recipient and time taken for self-care. These values could be manifested as a game with two objectives, one that encourages caregivers to express care (e.g., through virtual affirmations for time spent meeting the needs of the care recipient) and another one that encourages caregivers to be physically active (e.g., virtual rewards when PA goals are achieved). To accommodate temporal constraints, as opposed to exergames that require longer, sustained play sessions~\cite{agmon2011pilot,chao2015effects,brox2011exergames}, we suggest that exergames for AD caregivers should enable quick and intermittent game play (e.g., as is possible in \textit{casual games}). 

For example, an exergame could embed the value of spending time away from the game by giving bonus points for less interaction with the game. Designers typically seek to encourage increased user interactions with software that promotes wellness, as is exemplified by health technology research in which ``engagement'' has been defined as low or high based on the frequency with which users interact with the app~\cite{partridge2015effectiveness,guertler2015engagement}. For example, Guertler et al. defined engagement as ``the duration and frequency of involvement'' with a software tool that promoted increased physical activity~\cite{guertler2015engagement}. However, AD caregivers are one population in which reduced interactions are often necessary and valued to ensure that their loved one is receiving the care that he or she needs. We thus call for research that examines how to maximize the usefulness of exergames for caregivers, while minimizing the required interactions with such tools. 

Such a research agenda should build upon prior work exploring the design of lightweight interactions with health technologies. For example, research has explored the value of ``unlock journaling'' (i.e., quick ways of logging behavioral data) and ``glanceable displays'' (i.e., data visualizations through which information about behavioral progress can be quickly gleaned)~\cite{zhang2016examining, consolvo2008flowers}. However, more empirical research is needed to systematically examine what levels of interaction with exergames are needed to support various health outcome goals, such as increased physical activity levels or a greater sense of social connectedness. Different outcome goals likely bring with them different levels of health technology interaction needed to achieve the outcome. 

\subsection{Supporting Low-Burden Social Interactions}
Our storyboards incorporated story sharing as a proposed design approach to helping caregivers connect around and reflect upon their PA goal pursuits and their caregiving experience more generally. Our participants saw value in this type of story-sharing, both as a means of providing support to other AD caregivers and as a way to help non-caregivers better understand how AD affects the care recipient and caregiver. While existing commercial PA tracking tools enable users to share their PA data with others (e.g., through automated posts regarding goal attainment on social media platforms), our findings suggest that beyond data sharing, helping caregivers to craft and share stories around their PA pursuits may be a valuable direction for future work. 

However, our participants conveyed the importance of having social interactions that were low-burden, given how overburdened they already are with their caregiving duties. Indeed, research is needed to examine what types of social features are more and less effective at providing a sense of connectedness while avoiding the unintended consequence of placing increased stress on caregivers to interact with other system users. While some research has examined how online communities can support information sharing amongst caregivers~\cite{pagan2014use, liu2011improving}, open questions remain as to how systems can be designed to encourage participation in such sharing while mitigating feelings of additional burden. 

One approach may be to create storytelling scaffolds that help ease the process of sharing one's experiences with the caregiver community. For example, within the context of a social exergame, the tool could create composite multimedia stories that combine automatically-generated, abstract visualizations of the caregiver's progress towards their goal and interactive story building blocks that allow caregivers to easily author and share their experiences in more detail. These building blocks might be in the form of templates that help caregivers to quickly consider what information they might want to share with others, or auto-generated text suggestions that caregivers can select from and modify. In pursuing this design direction, an important research question would be to what extent stories created through pure user authoring, automation, and \textit{mixed initiative} interaction (i.e., via collaboration between the user and system)~\cite{horvitz1999principles} compare in terms of their impact on the creators and consumers of the content. One question for future work is, are stories created solely by the caregiver viewed as more or less authentic, informative, inspirational, and emotionally stirring than those created through automation and mixed-initiative interactions, and why? 

\subsection{Designing for Strategic Connection}
While our participants expressed an interest in hearing the stories of other caregivers, and engaging with them more generally within an exergame context, some participants also wanted to strategically connect with caregivers who have shared characteristics (e.g., a person who is caring for someone with the same AD stage, or who is also caring for a grandparent). This shared context was critical for helping our participants to feel that they could receive valuable support. As such, future work should explore how exergames can support social play, but specifically amongst caregivers with shared traits. For example, research could explore the benefits of a system that allows users to select the caregiver characteristics most important to them, and then matches caregivers with others who have those attributes. One challenge here would be helping users to avoid narrowing their field of social engagement prematurely, and unintentionally closing themselves off to the potential benefits of connecting with others they perceive to be too different from them. Indeed, connecting to a heterogeneous network of caregivers is important for the facilitation of bridging social capital, which can offer many important health benefits including access to novel information that might support them in their role as a caregiver and in their pursuit of their wellness goals~\cite{coriel2009social}. 

While much work has explored the benefits of online health communities and social health applications for helping people exchange support with individuals outside of their existing networks (e.g., through participation in online communities with strangers)~\cite{pagan2014use, maloney2002meaning, maloney2005multilevel, das2014influences}, there is little work exploring techniques that can help users strategically connect with people within this expanded network. For example, in an online health community designed for Spanish-speaking dementia caregivers, users could interact with other caregivers through posts on the site, but the site did not help caregivers navigate through the many caregivers using the site, to identify who might be most beneficial for them to connect with~\cite{pagan2014use}. Instead, the site simply provides a platform for interaction, with the assumption that caregivers will find their way to one another. This design approach is reflective of much existing social health technology that is designed to help people connect with individuals outside of their existing networks~\cite{wasilewski2017web}. To be sure, these communities have been shown to be useful environments for social support~\cite{wasilewski2017web}. However, our caregivers' desire to interact with others who have particular characteristics suggests that a fruitful research agenda will be one that examines the value of additional system scaffolding for connecting social exergame users to particular caregivers.
The desire to connect to those in similar situations has been found in other caregiver populations. In their work on caregivers for depressed family members, Yamashita et al. discuss the how the need to protect the privacy of the care recipient impedes a caregivers ability to seek support from other caregivers~\cite{yamashita2013understanding}.
Indeed, it may be challenging for people to, on their own, identify who in an exergame community might offer the greatest opportunities for informational, emotional, tangible, or appraisal support, especially while maintaining caregiver and care recipient control over their level of anonymity. Future work is needed to design and evaluate scaffolds that help caregivers navigate the sea of potential support partners within a social exergame context, particularly given that caregivers may have very limited time to do such social labor on their own. 

\subsection{The Felt Experience of Behavior Change}
Thus far, much of our discussion has centered on how technology can be designed to provide supports for caregivers whilst not introducing additional burdens given their already challenging lives. Continuing this thread of discussion, our work highlights the need for exergame research that attends to the emotional experience of caregiving and how that experience may impede or motivate behavioral change, as well as the emotional ramifications of taking time for oneself in the midst of caregiving. Indeed, some of our caregivers expressed a sense of guilt when taking time out for themselves and often wished for more hours in the day to balance the the numerous roles they filled. Our findings echo previous work on caregiver 'role strain,' the emotional burden that caregivers face when responsibilities to the care recipient inevitably conflict with the need for self-care, or even with non-caregiver responsibilities at home or work~\cite{bastawrous2013caregiver, chen2013caring}. Even when recognizing the benefits of physical activity and social connectedness of an exergame, our caregivers resisted exergame designs that would distract from their role as caregivers and create feelings of guilt. Thus, we encourage future research that studies how systems that attempt to encourage behavioral change impact caregivers emotionally, and how these systems can be leveraged to alleviate caregiver experience of role strain. For example, while an exergame that directs the caregivers' attention towards PA goal attainment may be beneficial for their physical health, there may be associated guilt with taking time out for oneself or negative feelings when caregiving responsibilities prevent a caregiver from achieving the PA goal. Research is needed that explores how technology can mitigate such feelings of guilt that arise from a caregiver's competing roles.

In addition, caregiving for a loved one with AD can be incredibly emotionally challenging, for example, as one watches a loved one decline cognitively and physically. As such, while within a social exergame physical activity and connectedness may be primary goals, providing features that promote positive mental and emotional health in caregivers will be an important focus area as well. Exergames are typically evaluated in terms of their impact on activity levels~\cite{peng2013using,kari2014can}. However, given that the emotional experience of caregiving is inseparable from caregivers' identity~\cite{chen2013caring}, an additional important outcome measure is how engagement with exergames impacts caregivers' emotional wellbeing.

\section{Conclusion}
We conducted semi-structured interviews and participatory design sessions with 14 AD informal caregivers to family members with AD. Our findings highlight the complex nature of caregiving, how this role profoundly impacts one's access to physical activity and social connections, and how this role influences caregivers' preferences towards exergame design. We further describe unique design considerations for how pervasive, social exergames might best support this population. Our findings highlight implications for design, including new opportunities for technologies that encourage PA and social connectedness promotion in a population that experiences significant barriers to wellness.

\section{Acknowledgements}
This work was supported by the Roybal Center Grant P30 AG048785. We thank our reviewers and colleagues for their feedback and suggestions.

% The next two lines define the bibliography style to be used, and the bibliography file.
\bibliographystyle{ACM-Reference-Format}
\bibliography{sample-base}

\end{document}